\documentclass[a4paper,11pt]{article}
\usepackage{pos}
\usepackage{euclid}

\title{\Euclid in a nutshell}

\renewcommand{\logo}{\relax}

\author*[a,b]{Antonino Troja}
\author[c,d]{Isaac Tutusaus}
\author[e,f,g]{Jenny G. Sorce}
\author{on behalf of the Euclid Consortium}

\affiliation[a]{Universit\`a degli Studi di Padova,\\
  Via Marzolo 8, Padova, Italy}
\affiliation[b]{INFN-PD,\\
  Via Marzolo 8, Padova, Italy}
  
\affiliation[c]{Institut de Recherche en Astrophysique et Plan\'etologie (IRAP), Universit\'e de Toulouse, CNRS, UPS, CNES,\\ 
14 Av. Edouard Belin, F-31400 Toulouse, France}
\affiliation[d]{Universit\'e de Gen\`eve, D\'epartement de Physique Th\'eorique and Centre for Astroparticle Physics,\\ 
24 quai Ernest-Ansermet, CH-1211 Gen\`eve 4, Switzerland}

\affiliation[e]{Univ. Lille, CNRS, Centrale Lille, UMR 9189 CRIStAL,\\
F-59000 Lille, France}
\affiliation[f]{Université Paris-Saclay, CNRS, Institut d’Astrophysique Spatiale,\\
91405, Orsay, France}
\affiliation[g]{Leibniz-Institut für Astrophysik (AIP),\\
An der Sternwarte 16, D-14482 Potsdam, Germany}

\emailAdd{antonino.troja@pd.infn.it}
\emailAdd{jsorce@aip.de}
\emailAdd{isaac.tutusaus@irap.omp.eu}

\abstract{\Euclid is a European Space Agency (ESA) mission designed to constrain the properties of dark energy and gravity via weak gravitational lensing and galaxy clustering. It will carry out a wide area imaging and spectroscopy survey 
in visible and near-infrared bands, covering approximately \num{15\,000} $\deg^2$ of the extragalactic sky in six years. \Euclid will be equipped with a 1.2 m diameter Silicon Carbide (SiC) mirror telescope 
feeding two instruments built by the Euclid Consortium: a high-quality panoramic visible imager 
and a near-infrared 
photometer 
and 
spectrograph
. These proceedings briefly describe the satellite and its instruments, which are optimised for pristine point spread function and reduced stray light, producing very crisp images. Furthermore, we summarise the survey strategy, the global scheduling, and the preparations for the satellite commissioning and the Science Data Centers to produce scientific data.}

\FullConference{%
  41st International Conference on High energy physics - ICHEP2022\\
  6-13 July, 2022\\
  Bologna, Italy
}

\begin{document}
\maketitle

\section{The \Euclid mission}

The \Euclid mission \cite{RedBook} is a survey aiming at describing the nature and the physics of the dark components of the Universe that drive its dynamics: dark matter and dark energy.  To do so, the satellite will measure billions of galaxy shapes and tens of millions of galaxy spectra from a privileged position: the Lagrangian point L2 in the two-body system Sun-Earth. The huge amount of data collected during the 6 years lifetime of the mission will permit us not only to go deeper in the understanding of the Universe's  dark sector, fulfilling the requirement of the \textit{primary science} but also to expand the knowledge of many astrophysical fields, that will be gathered under the so-called \textit{legacy science}. Both of them are described in the two proceedings papers accompanying this one \cite{Tutusaus_2022, Sorce_2022}. 

In this document, the \Euclid mission is concisely described
, as well as the instruments on-board. In the last section, we give an overview of global scheduling of the mission.

\subsection{The space survey}

\begin{figure}[!h]
    \centering
    \includegraphics[width=0.9\textwidth]{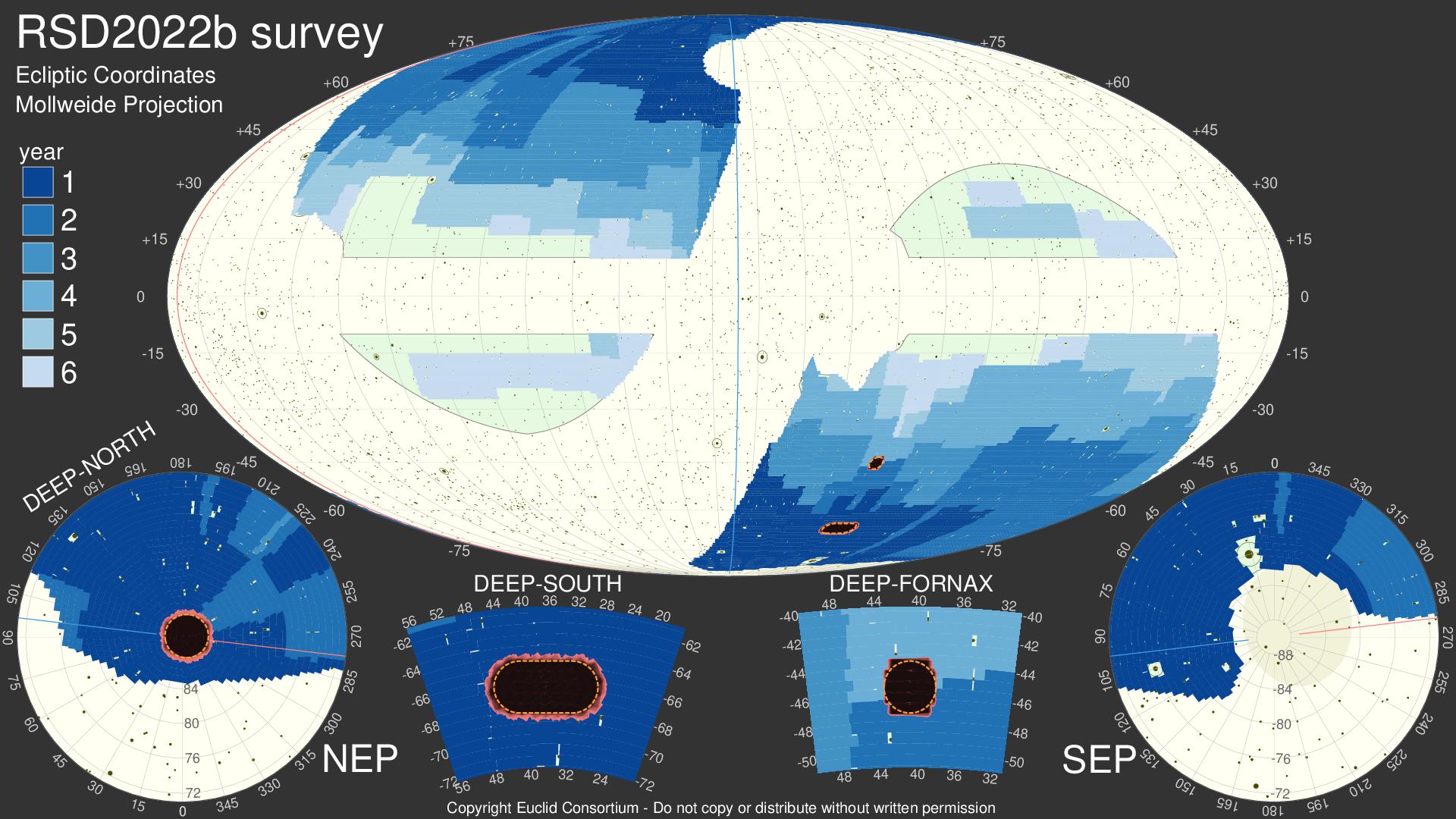}
    \caption{Euclid possible wide survey area is bounded by blue lines. The observed regions are coloured blue, the unobserved in green. The darker the blue, the earlier that patch of the sky is scanned. DS regions are shown below the Mollweide projection of the full sky, where the EWS area is depicted. 
    }
    \label{fig:SKY_AREA}
\end{figure}

The \Euclid spacecraft will observe the visible and near-infrared sky through a wide area imaging and spectroscopic survey, the Euclid Wide Survey (EWS, \cite{Scaramella_2022}). The EWS will cover approximately $15\,000 \deg^2$ during the nominal six-years mission lifetime (Fig. \ref{fig:SKY_AREA}). Together with the EWS, \Euclid will also perform an additional deep survey (DS).


The EWS is the core of the mission. It will allow us to detect the signatures of dark matter and dark energy, with the goal of better understanding the accelerated expansion of the Universe and the nature of dark energy. The two main probes of the EWS are 
galaxy clustering (GC) and weak lensing (WL).
GC 
will be measured from tens of millions of spectroscopic redshifts, in the range $0.7<z<1.8$ and an accuracy of $0.001 (1+z)$ at $68\%$ confidence level. WL will be measured using $\sim1.5$ billion galaxy shapes and will be coupled to photometric redshifts (photo-\textit{z}). Together
with complementary ground-based photometry, the near-infrared passbands enable the calculation of the mean photo-\textit{z} of the tomographic redshift bins with an accuracy of  $0.002\,(1+z)$ \cite{misha}.
WL measurements will be 
complemented by further cosmological probes derived from the same data and combined with external ones, e.g., clusters of galaxies, and cosmic microwave background (CMB) cross-correlation.

The EWS will cover more than one-third of the sky, 
excluding the Galactic and the ecliptic planes. The EWS is estimated to have an average limited magnitude of $26.5$ mag in the visible band \IE\, and $24.5$ mag in the near-infrared bands \YE, \JE, and \HE\,  ($5\sigma$ point-like source, magnitudes are in the AB magnitude system, \cite{Scaramella_2022}). On the spectroscopic side, the $H_\alpha$ line flux limit is $2\times10^{-16}\,\mbox{erg}^{-1}\,\mbox{cm}^{-2}\,\mbox{s}^{-1}$ at $1600$ nm.


%
%

The DS is designed to observe dimmer objects in smaller areas, covering at least $40\deg^2$ over three fields in the northern and southern Galactic hemispheres. 
DS will provide science calibration data sets to the EWS and data for legacy science, like faint high-redshift galaxies, quasars, and active galactic nuclei (AGN).

\subsubsection {The survey strategy}


Every day, \Euclid can observe a circle in ecliptic coordinates, whose size has a latitude varying width. 
The Sun-spacecraft axis will move $\sim1\degree$ per day, forcing the satellite to be as much as possible perpendicular to the Sun. In this way, the Sun-shield of the spacecraft will face the Sun at any time, maintaining the thermal stability of the mirrors and the instruments.

Each tile of the survey will be observed in `step-and-stare' mode, i.e., taking imaging and spectroscopic measurements after pointing the telescope to a fixed position of the sky before moving to the next one. The \Euclid field of view (FoV), common to the instruments, is about $0.56\deg^2$.
For each tile, \Euclid will follow a \textit{dithering technique}, consisting in taking a series of four observations, slightly changing the telescope pointing direction between them. This strategy aims at correcting in part instrumental effects on the images, for example, due to the gap between detectors in the focal planes.


\subsection{The ground segment}

\Euclid will collect more than half a million visual and near-infrared images during the 6 years of the mission. These images will be transferred to Earth daily. The data stream is expected to be approximately $100\,\rm{GB}$ per day, compressed by a factor of about $2.2$. The data taken by the satellite will be complemented by data from ground-based telescopes, covering the same sky as the EWS. The Science Ground Segment (SGS) will process and analyse the heterogeneous data sets composed of space and ground-based images. 
Furthermore, the SGS will be responsible for the \Euclid archives and the production of the official \Euclid data release.

\section{Instruments}

Two modules set up the \Euclid satellite: the Service Module (SVM) and the Payload Module (PLM). The SVM is under the responsibility of Thales Alenia Space (TAS), which is the prime contractor for the entire satellite. It comprises the instruments like the star tracker, gyroscopes, micro-motions systems, Attitude and Orbit Control Systems (AOCS), the thermal regulations system, 
the downlink communication system, 
tanks for hydrazine and cold gases, and the Fine Guidance Sensor (FGS).

The PLM 
is under the responsibility of Airbus (Defence and Space) and contains the telescope and the two main scientific instruments, the visual imager (VIS, \cite{Cropper_2022}) and the Near-Infrared Spectrometer and Photometer (NISP, \cite{Medinaceli_2022}), in addition to the PLM thermal control system. 
VIS and NISP were both delivered by the Euclid Consortium. 


\subsection{The telescope}

The telescope is an off-axis 3-mirror Korsch cold telescope. The primary mirror has a diameter of 1.2 meter, with a 
$\sim0.91\deg^2$ FoV. During the whole operative life of the survey, the main mirror will be cooled and maintained at a temperature of $T<130\,\rm K$. The thermal stability is assessed at around $50\,\rm mK$. The three mirrors are built in Silicon Carbide (SiC), a material that ensures excellent thermo-elasticity and stiffness. Furthermore, it is immune to ionizing radiations, guaranteeing high-performance when in space. The secondary mirror is provided with a 5 degrees-of-freedom mechanism (M2M) that allows movements of the mirror of the order of $1\micron$, allowing in-flight operations aimed to correct focus and tilt.

\begin{figure}[!t]
    \centering
    \includegraphics[scale=0.9]{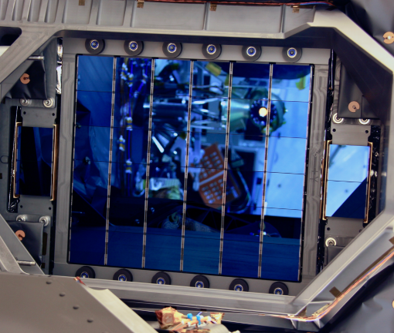}
    \includegraphics[scale=0.9]{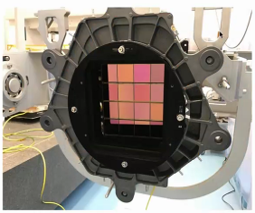}
    \caption{VIS (left) and NISP (right) focal plane. Courtesy of TAS.}
    \label{fig:VIS_NISP}
\end{figure}

\subsection{VIS}

VIS is the key to obtaining WL measurements. VIS  will measure the shapes of galaxies 
in the visible spectrum, covering the 550 to 900 nm wavelength region using a broad-band filter. An array of $6\times6$ e2V CCDs is mounted on the VIS focal plane (Fig. \ref{fig:VIS_NISP}), equivalent to about $0.56\deg^2$ FoV. 
Each CCD is made of $4096\times4132$ $12\mu$m $\ang{;;0.1}$ pixels. The image quality is expected to be around $\ang{;;0.23}$, allowing measurements of galaxies shapes with high-accuracy. 

\subsection{NISP}

\begin{figure}[!ht]
    \centering
    \includegraphics[width=0.9\textwidth]{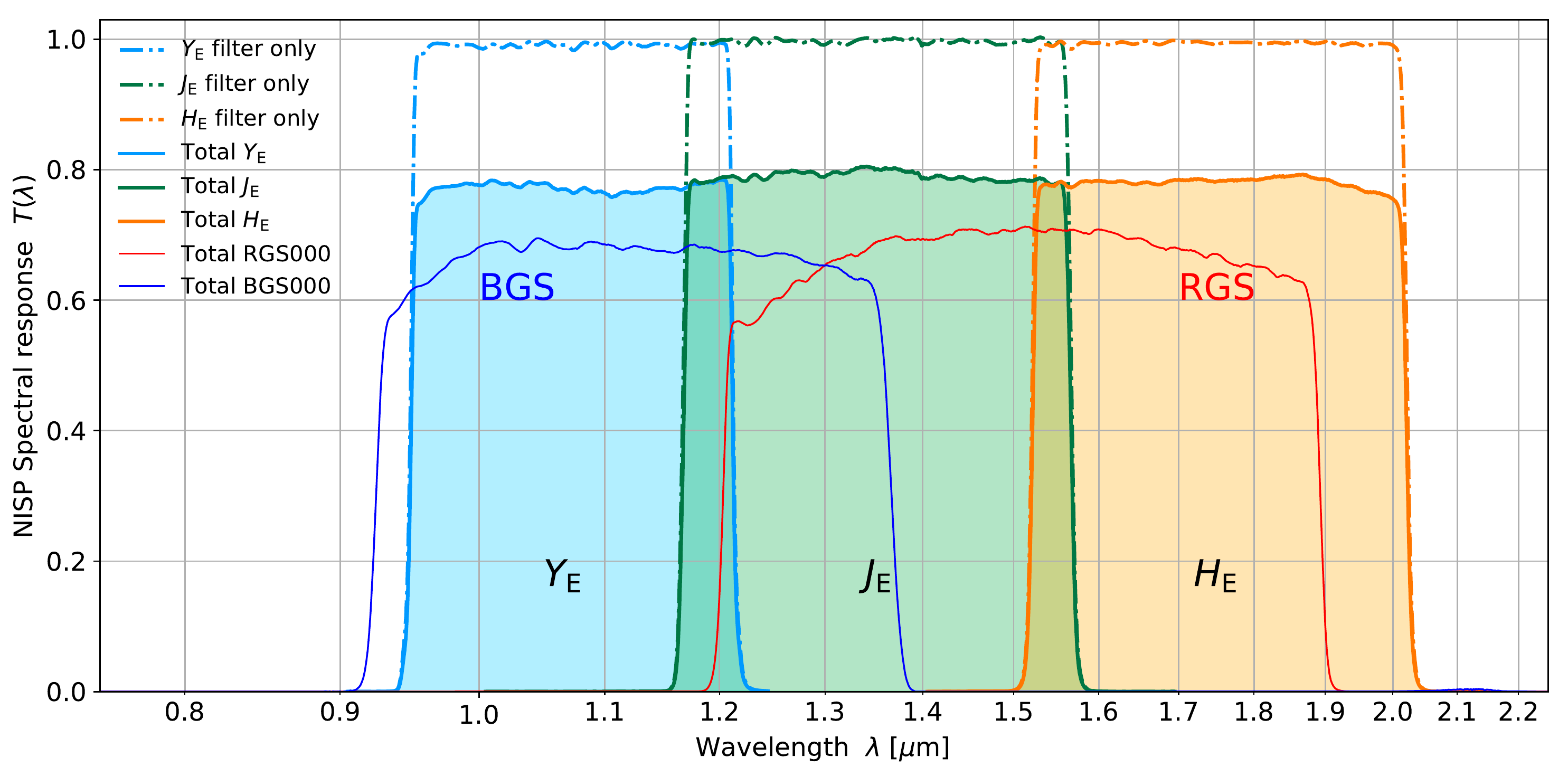}
    \caption{Transmission curves of the $Y_{\text E}$, $J_{\text E}$, and $H_{\text E}$ filters and the blue (BGS) and red grims (RGS). Figure based on \cite{misha}}
    \label{fig:NISP WL}
\end{figure}

NISP will observe the sky in the near-infrared spectrum (wavelength coverage of 900--2020 nm, Fig. \ref{fig:NISP WL}), acquiring photometric and spectroscopic data of the extra-galactic sources
in the FoV. The NISP focal plane (Fig. \ref{fig:VIS_NISP}) is equipped with an array of $4\times4$ HAWAII-H2RG detectors. Each detector is composed of $2048\times2048$ $18\mu$m $\ang{;;0.3}$ pixels and the full FoV is as large as the VIS one.

Near-infrared photometry will be performed on the same sources whose image is collected by VIS. 
Three filters will be used for photometry, $Y_\text{E}$, $J_\text{E}$, and $H_\text{E}$, covering wavelengths in the range $950-2020$ nm 
(Fig. \ref{fig:NISP WL}). The three filters are hosted on the Filter Wheel (FWA), together with two other slots: OPEN, used to let the light pass without filtering, to get its spectrum through the Grism Wheel (GWA), and CLOSED, to measure dark frames, useful to estimate the thermal response of the detectors.

NISP will get the spectra of all the sources in the FoV through slitless spectroscopy.
The light coming from the telescope will be dispersed through one of the grisms mounted on the GWA 
(namely RGS000, RGS180, and RGS270), 
in the wavelength range between 1200 and 1850 nm. Each grism will give the spectra along a different direction. These spectra will be combined offline in order to purify the spectrum of each source from contaminants. 
The grism RGS270 will not be used because of non-conformity
discovered in 2020 \cite{Laureijs_2020}. The observations with RGS000 and RGS180 will be complemented by two observations made again with RGS000 and RGS180, tilted by $-4\degree$ and $+4\degree$, respectively.
A fourth grism (BGS000) will cover the range between 920 and 1400 nm, and it will be used only in DS. 

\section{Global scheduling}

The \Euclid spacecraft launch is scheduled for Q3 2023. \Euclid will reach L2 in about 1 month. During the transfer phase to the target orbit, a 1-month-long commissioning phase will take place. After commissioning, a performance verification (PV) phase will be executed for 2 months, before starting the actual data collection. Three main data releases are expected. The first one (DR1) 24 months from the beginning of the survey, with data from about $2500\deg^2$ of the sky. The second data release  (DR2) is expected 2 years after DR1, covering about $7500\deg^2$. The third data release (DR3) will be delivered after 3 further years from DR2, with the full EWS area observed \cite{Scaramella_2022}.


\acknowledgments
IT acknowledges funding from the Swiss National Science Foundation and from
the European Research Council (ERC) under the European Union’s Horizon 2020 research and innovation
program (Grant agreement No. 863929). JS acknowledges support from the French Agence Nationale de
la Recherche for the LOCALIZATION project under grant agreements ANR-21-CE31-0019. The Euclid
Consortium acknowledges the ESA and a number of agencies and institutes that have
supported the development of \Euclid, in particular the Academy of Finland, the Agenzia Spaziale Italiana,
the Belgian Science Policy, the Canadian Euclid Consortium, the French Centre National d’Etudes Spatiales,
the Deutsches Zentrum für Luft- und Raumfahrt, the Danish Space Research Institute, the Fundação para
a Ciência e a Tecnologia, the Ministerio de Ciencia e Innovación, the National Aeronautics and Space
Administration, the National Astronomical Observatory of Japan, the Netherlandse Onderzoekschool Voor
Astronomie, the Norwegian Space Agency, the Romanian Space Agency, the State Secretariat for Education,
Research and Innovation (SERI) at the Swiss Space Office (SSO), and the United Kingdom Space Agency.
A complete and detailed list is available on the Euclid web site (http://www.euclid-ec.org).

\end{document}